\documentclass[11pt,a4paper]{article} 
\usepackage{epsfig} %include package for eps-files

\title{Baby Skyrme models for a class of potentials}  \author{Parvin Eslami
\footnote{Permanent address: Dept. of Physics, University of Mashhad, Iran}\\
Wojtek Zakrzewski\footnote{email: W.J.Zakrzewski@durham.ac.uk}\\
\\Centre for Particle Theory,\\Mathematical Sciences,  \\ University of
Durham\\
\\
Mohsen Sarbishaei\\
Dept of Physics, University of Mashhad, Iran\\}

\newcommand{\ph}{\vec{\phi}} \newcommand{\pha}{\phi_{a}}
\newcommand{\dmu}{\partial_{\mu}} \newcommand{\umu}{\partial^{\mu}}
\newcommand{\dnu}{\partial_{\nu}} \newcommand{\unu}{\partial^{\nu}}
\newcommand{\di}{\partial_{i}} 
\renewcommand{\dj}{\partial_{j}}  \hoffset =
-2cm \textwidth = 170mm

\begin{document}

\maketitle 

%%%%%%%%%
\abstract

{We consider a class of (2+1) dimensional baby Skyrme models with 
potentials that have more than one vacum.
These potentials are generalisations of old and new baby Skyrme models;
they involve more complicated dependence on $\phi_3$.
We find that when the potential is invariant under $ \phi_3
\longrightarrow -\phi_3$ the configurations 
corresponding to the baby skyrmions lying ``on top of each other" 
are the minima of the energy.
However, when the potential breaks this symmetry the lowest field
configurations correspond to separated baby skyrmions.
We compute  the energy distributions for skyrmions 
of degrees between one and eight and discuss their geometrical shapes 
and binding energies.
We also compare the 2-skyrmion states for these potentials.
Most of our work has been performed numerically with the model 
being formulated in terms of three real scalar fields
(satisfying one constraint).}

%%%%%%%%%
\section{Introduction}

The Skyrme model describes a non-linear  theory for SU(2) valued fields  
which has soliton solutions.\cite{sky:61} 
Although the potential term is optional in the (3+1)
dimensional nuclear Skyrme model 
its presence is necessary in (2+1) dimensions to ensure the stability
of these solitonic solutions.
In this paper we discuss multisolitons in a two-dimensional
version of the  model for a class of potentials which generalises
the previously studied cases.
In the literature we can find three specific
potentials that have been studied in 
some detail.
%  (see \cite{pz:93},\cite{psz:95a},\cite{weidig:99}).

The holomorphic model has only one stable solution 
(describing one skyrmion) and this solution has a simple 
analytical form \cite{pz:93}.
The old baby Skyrme model \footnote{this name was introduced in \cite{weidig:99}}
has stable solutions
for any number of skyrmions. In \cite{psz:95a} they
were shown to lead to a crystalline lattice of skyrmions.
The new baby Skyrme model was studied in detail in \cite{weidig:99}
where it was shown that its solutions corresponded to field configurations
with radially symmetrical energy densities - which correspond
to many skyrmions lying ``on top of each other".
We generalise these studies by considering
potentials with a more general dependence on $\phi_3$ thus 
leading to models with more vacua. 

%%%%%%%%%%%%%%%%%%
\section{A Skyrme model for a class of potentials}

The Skyrme model was first proposed by T.H.R Skyrme 
\cite{sky:61} in 1960.
Its classical solutions fall into various classes
characterised by a topological number to be identified
with the baryon number. Thus the model can describe
mesons as well as various baryon configurations. 
The baby Skyrme models \cite{pz:93} are (2+1) dimensional versions 
of the Skyrme model and its  topologically nontrivial solutions  
are called baby skyrmions.

The Lagrangian density of the model contains three terms, from left to right:
the pure $S\sp2$ sigma model, the Skyrme and the potential terms:

\begin{equation}\label{lagrangian} 
{\cal L}=\partial_{\mu}\ph\cdot\partial^{\mu}\ph
-\theta_{S}\left[(\partial_{\mu}\vec{\phi} \cdot
\partial^{\mu}\vec{\phi})^{2} -(\partial_{\mu}\vec{\phi} \cdot
\partial_{\nu}\vec{\phi}) (\partial^{\mu}\vec{\phi} \cdot
\partial^{\nu}\vec{\phi})\right] - \theta_{V} V(\ph).
\end{equation}

The vector $\ph$ is restricted to lie on a unit sphere ${\cal S}^{2}$
hence $\ph\cdot\ph=1$.\\

Note that to have finite potential energy the field at spatial infinity 
cannot depend on the polar angle $\theta.$  

\begin{equation}
\label{BCinf} \lim_{r \rightarrow\infty} \ph(r,\theta)= {\ph}^{(0)}.  
\end{equation}
Hence this boundary condition defines an one-point compatification of $R_2$,
allowing us to consider $\ph$ on the extended plane 
$R_2\bigcup{\infty}$  topologically equivalent to  ${\cal S}^{2}$.
In consequence, the field configurations are maps   
\begin{equation} 
{\cal{M}}: {\cal{S}}^{2} \longrightarrow {\cal S}^{2}.  
\end{equation}
which can be labeled by an integer valued topological index $Q$:
\begin{equation} 
Q=\frac{1}{4\pi}\epsilon^{abc}\int dxdy\phi_{a}
\left(\partial_{x}\phi_{b}\right) \left(\partial_{y}\phi_{c}\right).
\end{equation} 
As a result of this non-trivial mapping the model has 
topologically nontrivial solutions 
which describe ``extended structures", namely, baby skyrmions . 
The different choices of the potential term lead to various shapes
of the energy density of these baby skyrmions.

The equation of motion for a general potential that depends 
on $\phi_{3}$ takes the form:
\begin{center}
\begin{eqnarray}
  \dmu\umu\pha-(\ph\cdot\dmu\umu\phi)\pha -2\theta_{S}[ (\dnu\ph\cdot\unu\ph)
  \dmu\umu\pha +(\dmu\unu\ph\cdot\umu\ph)\dnu\pha \nonumber \\
  -(\unu\ph\cdot\umu\ph)\dnu\dmu\pha -(\dnu\unu\ph\cdot\umu\ph)\dmu\pha
  +(\dmu\ph\cdot\umu\ph)(\dnu\ph\cdot\unu\ph)\pha \nonumber \\
  -(\dnu\ph\cdot\dmu\ph)(\unu\ph\cdot\umu\ph)\pha]
  +\frac{1}{2}\theta_{V}\frac{dV}{d\phi_{3}}(\delta_{a3}-\phi_{a}\phi_{3})=0.
\end{eqnarray}
\end{center}

It is convenient to rewrite this equation as:
\begin{equation}
\label{eom} \partial_{tt}{\phi}_{a}=K_{ab}^{-1} {\cal
  F}_b\left(\ph,\partial_{t}\ph,\partial_{i}\ph\right) 
\end{equation}
with
\begin{equation}
K_{ab}=(1+2\theta_{S}\partial_{i}\ph\cdot\partial_{i}\ph)\delta_{ab}
-2\theta_{S}\partial_{i}\phi_a\partial_{i}\phi_b 
\end{equation}
and a rather complicated expression for ${\cal F}_b$.

Then to find a solution to this equation  we invert the matrix $K$ and 
simulate the time evolution by a 4th order Runge Kutta method
supplemented by the imposition of a correction due 
to the constraint $\ph\cdot \ph=1$.

The kinetic  and  potential energy
densities of the baby Skyrme model are:
\begin{equation}  
 {\cal K}=\left(\partial_{t}\ph\cdot\partial_{t}\ph\right) 
\left(1+2\theta_{S}\left(\di\ph\cdot\di\ph\right)\right)
-2\theta_{S}\left(\partial_{t}\ph\cdot\di\ph\right)^{2}
\end{equation}

\begin{equation}  
 {\cal V}=\left(\di\ph\cdot\di\ph\right)+ \theta_{S}\left[
\left(\di\ph\cdot\di\ph\right)^{2}-
\left(\di\ph\cdot\dj\ph\right)\left(\di\ph\cdot\dj\ph\right) \right]
+\theta_{V}V(\phi_3).  
\end{equation}

The potentials we want to consider in this paper are of the form
\begin{equation}
V(\phi_3)=(1-\phi_3)G(\phi_3),
\end{equation}
where $G(\phi_3)=1$ or $(1+\phi_3)$ or $\phi_3\sp2$ or
$(1+\phi_3)\phi_3\sp2$. Thus this potential vanishes at $\phi_3=1$
and also at $\phi_3=-1$ and/or $\phi_3=0$.

The total energy takes the form:
$$E=\int dxdy \left({\cal V}+{\cal K}\right).$$

%%%%%%%%%%%%%
\section{Static Solutions}

In this paper we are primarily interested in finding minimal
energy configurations corresponding to many skyrmions. We want 
to see, for different forms of the potential, what such configurations 
are like and what properties
they have.

We will consider a numerically 
found configuration to be a multiskyrmion if it is satisfies certain   
numerical checks for a local minimum and if, moreover, its energy satisfies

$E_n<E_k+E_l$ \quad \mbox{for all integers}$\quad 1<$ $\ell$ \mbox{and} $k<n$ \quad \mbox{such
 that}  $k+\ell=n$

We have included this condition in our definition because we want 
multisoliton to be stable with respect to decay into multiskyrmions 
of smaller degree.

There are many ways of finding static solutions.
First, as the potential is only a function of $\phi_3$, there
is a symmetry corresponding to rotations around the $\phi_3$ axis. So
choosing the spatial dependence conveniently it is clear that
fields which correspond to a generalised multihedgehog ansatz
will be static solutions of the equations of motion \cite{psz:95a}.
Such fields are given by
{\it ie}
\begin{equation} \label{hedgehog} 
\vec{\phi}=\left(\matrix{\sin[f(r)]\cos(n\theta-\chi)\cr\sin[f(r)]
\sin(n\theta-\chi)\cr\cos[f(r)]\cr}\right).   
\end{equation}% 
where (r,$\theta$) are polar coordinates in the $x-y$-plane 
and the function $f(r)$, called the profile function, is required to
satisfy certain boundary conditions to be specified below.
The angle $\chi$ is arbitrary, but fields with different 
$\chi$ are related by an iso-rotation and are therefore degenerate in energy.
$n$ is a non-zero integer and equals to the topological charge.
   
For the $\phi$ field be regular at the origin 
the profile function $f(r)$ has to be satisfy :
\begin{equation}
\label{BC0} f(0)=m\pi,       
\end{equation}
where $m$ is an integer.

We choose 
the vacuum at infinity to be ${\ph}^{0}=(0,0,1)$  and this results
in another boundary condition, namely:
\begin{equation}
\label{BCinf} \lim_{r \rightarrow\infty} f(r)=0.  
\end{equation}

The total energy of the field configuration then takes the form:
\begin{equation}
\label{lag}
E=(4\pi)\frac{1}{2}\int_{0}^{\infty} rdr\left( {f^{'}}^{2} +
n^{2}\frac{\sin^{2}f}{r^{2}}(1+2\theta_{S}{f^{'}}^{2}) +
\theta_{V}\mbox{\~{V}}(f)\right),   
\end{equation}
where $f^{'}={df\over dr}$ and $\mbox{\~{V}}(f)=V(\phi_3)$. To determine the profile function $f(r
)$ we treat (\ref{lag}) as a
 variational problem and we get a second-order ODE for $f$:
\begin{eqnarray}
\label{profile}  \left(r+\frac{2\theta_{S}n^{2}\sin^{2}f}{r}\right)f^{''} +
  \left(1-\frac{2\theta_{S}n^{2}\sin^{2}f}{r^{2}}+
  \frac{2\theta_{S}n^{2}\sin{f}\cos{f} f{'}}{r} \right)f^{'} \nonumber \\
  -\frac{n^{2}\sin f\cos f}{r}
  -r\frac{\theta_{V}}{2}\frac{d\mbox{\~{V}}(f)}{df} =0. 
\end{eqnarray} 
This equation then has to be solved numerically ({\it ie}
via the shooting method).

The topological charge takes the form:
\begin{equation}
Q=-\frac{n}{2}\int_0^{\infty}r dr\left(\frac{f^{'}\sin f}{r}\right)=
 \frac{n}{2}[\cos f(\infty)-\cos f(0)].  
\end{equation} 

This equation shows that to have an integer value for the topological 
charge, $m$ in (\ref{BC0}) must be an odd number.
In this paper we consider the solutions with $m=1$.

The behaviour of solutions of this equation near the origin and for 
large $r$ can be deduced analytically and was also discussed in
\cite{psz:95a}. 
The result is that
\begin{itemize} 
\item For small $r$, the profile function behaves as
\begin{equation} \label{f_origin} 
f \simeq \pi+C_{n}r^{n}
\end{equation}
and so
\begin{equation} \label{fp_origin} 
f^{'} \simeq nC_{n}r^{n-1} 
\end{equation}
 as long as $\frac{dV(f)}{df}$ tends to zero at this point. 
\item At large r, the ODE reduces to
\begin{equation}
\label{larger} f^{''}+\frac{1}{r}f^{'}-\frac{n^2}{r^2}f
  -\frac{\theta_{V}}{2}\frac{d\mbox{\~V}(f)}{df} {\big \vert}_{\mbox{large
 r}}=0. 
\end{equation}
 Using the boundary
conditions, we see that if
 $\frac{1}{f}\frac{d\mbox{\~V}}{df}{\big\vert}_{\mbox{large r}}\rightarrow 1$
the profile function decreases exponentially
\begin{equation}
f(r)\longrightarrow\frac{1}{\sqrt{\theta_{V}r}}\exp(-\theta_{V} r).  
\end{equation}
\end{itemize}
This means that the potential localizes the skyrmion exponentially.

Of course we do not know whether such fields are the global minima of the
energy. To check this we can perform a numerical simulation of 
(\ref{eom})
taking our derived field configurations (\ref{hedgehog}) as 
initial conditions.
We perturb them a little and then evolve them according to (\ref{eom})
with an extra dissipative term - $\gamma \partial_t\phi_a$ added to the right 
hand side of (\ref{eom}). We vary our perturbation and look at the 
configurations and their energies that our fields finally settle at.
When the original configurations are the global minima of the energy
the perturbed fields evolved back to them; otherwise (for sufficiently
large perturbations) the fields evolve to other static solutions and,
hopefully, the global minima. Another possibility, in particular to check
whether these solutions are global minima, is to start with a more general field
configuration corresponding to $n$ skyrmions, say, $n$ skyrmions located
on a circle, and then evolve it using dissipative dynamics.
In this case the circular set-up of $n$ 1-skyrmions with relative 
iso-orientation $\delta\chi=\frac{2\pi}{n}$
was obtained by using the combination
of 1-skyrmions determined by the hedghog field (\ref{hedgehog})
and combined together as discussed in the appendix.

The actual outcome ({\it ie} which fields are the global minima) depends on 
the form of the potential. In the next section we discuss the potentials that
we have used in our simulations.

%%%%%%%%%%%%%
\section{Different potentials}
\subsection{General comments}
A skyrmion is a $S\sp2\rightarrow S\sp2$ map and so as the field at spatial
$\infty$ corresponds to $\phi_3=+1$ ({\it ie} the ``North pole"
of the field $S\sp2$) and the skyrmion's position
is at the point for which $\phi_3=-1$. If $V(\phi_3=-1)$ does not vanish
it costs energy for the field to take this value; hence in this case
we expect the skyrmions to repel when they are brought
``on top of each other". Thus the minimal energy
multiskyrmion configurations should be different when $V(\phi_3=-1)=0$
and $\ne0$. In fact, this was seen in the earlier studies; in the ``new
baby Skyrme" model ($V(\phi_3)=1-\phi_3\sp2$) the skyrmions lie
on ``top of each other" while for the ``old baby Skyrme" model
($V(\phi_3)=1-\phi_3$) they are separated from each other by finite
distances. We have repeated these simulations and have extended 
them to a larger number of skyrmions. Below, we present our results:

%%%%%%

\subsection{$V=(1+\phi_{3})^4$}
The first potential studied in the baby Skyrme model was
$V=(1+\phi_{3})^4$ (\cite{lpz:90},
\cite{sut:91}, \cite{pz:93}).
This potential was chosen because the
equations of motion have an analytic static solution
(for $n=1$) of the form $W=\lambda (x+iy)$, where $\lambda$ is
related to $\theta_{S}$ and $\theta_{V}$.
The resultant soliton is only polynomially localised.
Two such skyrmions repel each other
and scatter at 90 degrees when sent towards each other with 
sufficiant speed. The model has further solutions
corresponding to several solitons ``on top of each other";
but these solutions are unstable. When perturbed the solitons
separate and move away to infinity.
Thus this model have no stable multi-skyrmion solutions.
%%%%%%

\subsection{$V=1-\phi_{3}$}
This potential was studied in some detail in \cite{psz:95a} and
\cite{psz:95b}.
As the value of the potential at $r=0\;(\phi_{3}=-1)$ is $V(\phi_{3})=2$
we expect some repulsion of the skyrmions when they are close
together. Thus a field configuration of $n$ skyrmions
on top of each other is unstable. This was studied a little
in \cite{psz:95a} where it was shown that as a result of this repulsion
the minimal energy multi-skyrmion field configurations produce nice
lattice-like patterns. We have repeated these studies and extended
them further to larger number of skyrmions. As there appear to
be many solutions we have performed many simulations starting
with different initial conditions and different perturbations. 

Our results show a rather complicated pattern of minimal energy
field configurations. The results are presented in table 1.
We use the coefficients of \cite{psz:95a} $\it ie$ add a factor $\frac
{1}{2}$ to the sigma model term, $\theta_S=0.25$ and $\theta_V=0.1$.
 
All configurations are built out of 1, 2 and 3 skyrmions. 
In fig. 1 we show the energy densities 
of a pair and a triple.

%\begin{figure}
%\begin{center}
%\includegraphics[width=8cm]{old2.eps}
%\includegraphics[width=8cm]{old3.eps}
%\caption{ Energy density of 2-skyrmion and 3-skyrmion states 
%for $V=1-\phi_{3}$ }
%\end{center}
%\end{figure}

A pair 
of skyrmions is very bound, more bound than a triple, and the pairs
and triples are further bound for larger values of $n$. Thus
the binding per skyrmion of a configuration of $4$ and $5$ skyrmions
is less that of a single pair or a single triple, respectively.

The first more interesting case is of $6$ skyrmions which seems to have several bound
states. Of these, the state corresponding to 3 pairs is the most
bound, the other two (2 triples and 6 individual ones) are
at best only local minima. For 7 skyrmions the lowest energy
configuration corresponds to 2 pairs and a triple, which can be thought 
of a bound state of 4 and 3. The other state, corresponding to 
a triple sandwiched between two pairs has a higher energy and is at
most a local minimum.
In fig 2. we present energy densities of these two configurations.
%%% here we put ps. files of two solutions for 7.

%\begin{figure}
%\begin{center}
%\includegraphics[width=8cm]{old7a.eps}
%\includegraphics[width=8cm]{old7b.eps}
%\caption{ Energy density of two states of $n=7$  
%for $V=1-\phi_{3}$ }
%\end{center}
%\end{figure}

For higher $n$ the situation becomes even more complicated; the lowest
energy states always involve sets of pairs, but there seem to be other states
involving triples and even singles.

One question one can ask is whether the $n=2$ state involves 
two skyrmions ``on top of each other" or slightly displaced. When 
started from the original configuration of two skyrmions ``on top
of each other" slightly perturbed the system evolves a little, but
this evolution is comparable to the original perturbation. Thus
we believe the two skyrmions are slightly displaced, but this
displacement is almost infinitesimal.

\begin{table}
\begin{center}
\begin{tabular}{|c||c|c|c|c|}\hline
  Charge & Energy & Energy per skyrmion  & description
  \cr\hline 1 & 1.549 & 1.549 & - 
  \cr\hline 2 & 2.920 & 1.460 & a pair 
  \cr\hline 3 & 4.400 & 1.466 & a triple 
  \cr\hline 4 & 5.827 & 1.457 & two pairs 
  \cr\hline 5 & 7.321 & 1.464 & triple + pair 
  \cr\hline 6 & 8.781 & 1.464  & 6 individual \cr & 8.731&1.455 & three pairs   \cr&8.790 &1.465 & two triples 
  \cr\hline 7 & 10.248 & 
  1.464 & pair+triple+pair\cr & 10.230& 1.461& triple+2 pairs 
  \cr\hline 8 & 11.688& 1.461& 4 pairs
  \cr\hline 9 & 13.107 & 1.456 & 4 pairs + single 
  \cr\hline 10& 14.550 & 1.455& 5 pairs 
  \cr\hline  
\end{tabular}
\end{center}
\caption{Multi-skyrmions of the old baby Skyrme model
 with $V=1-\phi_{3}$ }
\end{table}

%%%%%%%    

\subsection{$V=1-\phi_{3}^{2}$}

The model with this potential was investigated
in great detail by Weidig \cite {weidig:99}.
As both $\phi_3=\pm1$ correspond to the vacuum
there is no energy argument which stops
the skyrmions from ``lying on top of each other".
This is in fact what Weidig  saw in his simulations; the states 
found by the shooting method (and so corresponding to
many skyrmions ``on top of each other") are the 
minima of the energy. Thus the energy densities 
of lowest energy multiskyrmion configurations have a ring-like
structure.

%%%%%%%

\subsection{$V=\phi_{3}^{2}(1-\phi_{3}^{2})$}
The extra factor $\phi_{3}^{2}$ in this potential
causes it to have an additional zero and so the vacuum structure
is even richer. 
However, like for the $1-\phi_3\sp2$ potential, there are no
energetic arguments which would disfavour the skyrmions
lying ``on top of each other".   
Indeed, this is what the simulations have shown; and we have checked
this by starting with the hedgehog ansatz and $n$ solitons
on a circle. The simulations have shown that
the hedghog ansatz 
configurations are the minimal-energy
solutions. The shapes of energy density have radial symmetry and resemble 
the configurations of Weidig. The difference is in the energy
levels (in this potential the energy per skyrmion decreases
monotonically with $n$). Our results are presented in table 
2. In this (and the next table)
we have added a factor $\frac{1}{2}$ to the sigma term and chose the 
$\theta$ coefficients to be $\theta_S=0.2$ and $\theta_V=0.05$.
The values of the energies are divided by $4\pi$. 

\begin{table}
\begin{center}
\begin{tabular}{|c||c|c|c|c|}\hline
  Charge & Energy & Energy per skyrmion & Break-up modes & Ionisation Energy
  \cr\hline 1 & 1.281 & 1.281 & - & - \cr\hline 2 & 2.332 & 1.166 & $1+1$ &
  0.231 \cr\hline 3 & 3.430 & 1.143 & $2+1$ & 0.184 \cr\hline 4 & 4.540 &
  1.135 & $2+2$ & 0.125 \cr & & & $3+1$ & 0.172 \cr\hline 5 & 5.654 & 1.131 &
  $3+2$ & 0.107 \cr & & & $4+1$ & 0.166 \cr\hline 6 & 6.771 & 1.129 & $3+3$ &
  0.087 \cr & & & $4+2$ & 0.099 \cr & & & $5+1$ & 0.164 \cr\hline 7 & 7.891 & 
  1.127 & $4+3$ & 0.078 \cr & & & $5+2$ & 0.095 \cr & & & $6+1$ & 0.162
 \cr\hline 8 & 9.011 & 1.126 & $4+4$ & 0.067 \cr & & & $5+3$ & 0.072 
 \cr & & & $6+2$ & 0.092 \cr & & & $7+1$ & 0.161 \cr\hline  
\end{tabular}
\end{center}
\caption{Multi-skyrmions of the baby Skyrme model
 with $V=\phi_{3}^{2}(1-\phi_{3}^{2})$ }
\end{table}

%%%%%%%

\subsection{$V=\phi_{3}^{2}(1-\phi_{3})$}

Like in the previous case the extra factor $\phi_{3}^{2}$ 
changes the nature of the solutions.  
In a way the effects due to this potential are
closer to those of the ``old baby Skyrme model", in that 
$V(\phi_3=-1)\ne0$.
 
We have found that for this potential only the hedghog ansatz with $n=1$
has minimal-energy and that for $ 1 < n \le 8 $ 
the skyrmions ``on top of each other", when
perturbed, evolve into configurations corresponding  to
 $n$-skyrmions that lie on regular polygons.
When starting from a circular set-up for this potential we 
have calculated the total 
energy of $n$ 1-skyrmions as a function of the radius of the circle to get the 
initial state with minimum energy. This way the initial state settles 
down to the stable state sooner. We have confirmed this
by performing also some simulations with different radi.
We have also added to the initial
configuration a non-symmetrical exponentially decaying
perturbation to check whether the resultant multi-soliton states are stable,
and whether, after dissipation, we reproduce the previous solutions.

%\begin{figure}
%\begin{center}
%\includegraphics[width=8cm]{p5_2.eps}
%\includegraphics[width=8cm]{p5_3.eps}
%\caption{ Energy density of  2-skyrmion and 3-skyrmion states for 
%$V=\phi_{3}^{2}(1-\phi_{3})$ }
%\end{center}
%\end{figure}

In fig. 3 we present picture of the energy density of the lowest
energy field configurations involving 6 skyrmions.

%\begin{figure}
%\begin{center}
%\includegraphics[width=8cm]{p5_4.eps}
%\includegraphics[width=8cm]{p5_6.eps}
%\caption{ Energy density of  6-skyrmion for 
%$V=\phi_{3}^{2}(1-\phi_{3})$ }
%\end{center}
%\end{figure}

Our results show that this time (when we compare this case 
with $V=1-\phi_3$) the bindings are more comparable; in fact 
three skyrmions are more bound than two, and so the pattern is very different.
It is interesting to note that the most bound system involves 6
skyrmions.

Our results on the binding energies {\it etc} are presented in table 3.

\begin{table}
\begin{center}
\begin{tabular}{|c||c|c|c|c|}\hline
  Charge & Energy & Energy per skyrmion & Break-up modes & Ionisation Energy
  \cr\hline 1 & 1.249 & 1.249 & - & - \cr\hline 2 & 2.372 & 1.186 & $1+1$ &
  0.127 \cr\hline 3 & 3.530 & 1.177 & $2+1$ & 0.091 \cr\hline 4 & 4.693 &
  1.173 & $2+2$ & 0.050 \cr & & & $3+1$ & 0.086 \cr\hline 5 & 5.859 & 1.172 &
  $3+2$ & 0.043 \cr & & & $4+1$ & 0.084 \cr\hline 6 & 6.926 & 1.154 & $3+3$ &
  0.133 \cr & & & $4+2$ & 0.138 \cr & & & $5+1$ & 0.182 \cr\hline 7 & 8.131 & 
  1.162 & $4+3$ & 0.092 \cr & & & $5+2$ & 0.100 \cr & & & $6+1$ & 0.045
 \cr\hline 8 & 9.300 & 1.163 & $4+4$ & 0.086 \cr & & & $5+3$ & 0.088 
 \cr & & & $6+2$ & 0.002 \cr & & & $7+1$ & 0.080 \cr\hline  
\end{tabular}
\end{center}
\caption{Multi-skyrmions of the baby Skyrme model
 with $V=\phi_{3}^{2}(1-\phi_{3})$  }
\end{table}

%%%%%%%

\section{Comparison of potentials between  2 skyrmions}

We have also looked at the ``potentials" between two skyrmions in our models.
To do this we had to decide how to construct field configurations 
involving two skyrmions.
We decided to do this as follows:
We have taken $f(r)$ for $n=1$ and computed the field $W$ from (\ref{wcom}). 
Then we combined two such fields with $\delta\chi=\pi$
and varied the distance between them computing the energy as a 
function of the distance. We examined the following ways of combining
 2 skyrmions:

\begin{equation}\label{one}
W=W_1+W_2
\end{equation}
\begin{equation}\label{two}
W=W_1-W_2
\end{equation}
\begin{equation}\label{three}
W=W_1+W_2-W_1W_2
\end{equation}
\begin{equation}
W=W_1-W_2+W_1W_2.
\end{equation}
As could be expected, we have found that for very small $r$ none of them  
gives reliable results.
However, when two skyrmions are well separated the combination (\ref{three})
gives us the lower energy than (\ref{one}) and so approximates
the 2 skyrmion field more accurately.
The figures presented below
show our numerical results for the following potentials 

\begin{equation}\label{a}
V=(1-\phi_3)
\end{equation}
\begin{equation}\label{b}
V=(1-\phi_3^{2})
\end{equation}
\begin{equation}\label{c}
V=\phi_3^{2}(1-\phi_3^{2})
\end{equation}
\begin{equation}\label{d}
V=\phi_3^{2}(1-\phi_3)
\end{equation}

As expected when the two skyrmions are far from each other the energy 
approaches 
twice the energy of one skyrmion. In our plots fig 5. we have indicated
the energy of the hedgehog field with $n=2$. Clearly, two
skyrmions of this field are at $r=0$ but for better visualisation we 
have indicated it 
by a line. 
The low $r$ part of each plot (with points indicated by $\triangle$) overestimates
the energy and cannot be trusted (at these points the skyrmions deform
each other and are not given by ansaetze  (21-24). However, as the
real energies are lower we see that our
results confirm again that for the potentials (\ref{b}) and 
(\ref{c}) the hedgehog fields are minimal-energy solutions
but for (\ref{a})  and (\ref{d}) the 
minimal-energy solutions are different. The difference is more obvious 
in the (\ref{d}) case as the plot indicates that the minimum
of the energy is at $r=3.75$.
For (\ref{a}) it may seem that the hedgehog field with $n=2$ is 
the minimal energy
state but when we have evolved it the energy density expanded a bit
thus showing that two baby skyrmions separate from each other a little.
At the same time the energy has decreased and so we can conclude that for 
this potential the minimum of energy field configuration 
also differs from the hedgehog field. Incidentally the $\theta's$
used to calculate the potentials (\ref{a}) and (\ref{b}) were chosen so that all four curves are compareable. Thus they are different 
from values used in table 1 and 4. 

\begin{figure}
\begin{center}
\includegraphics[width=8cm]{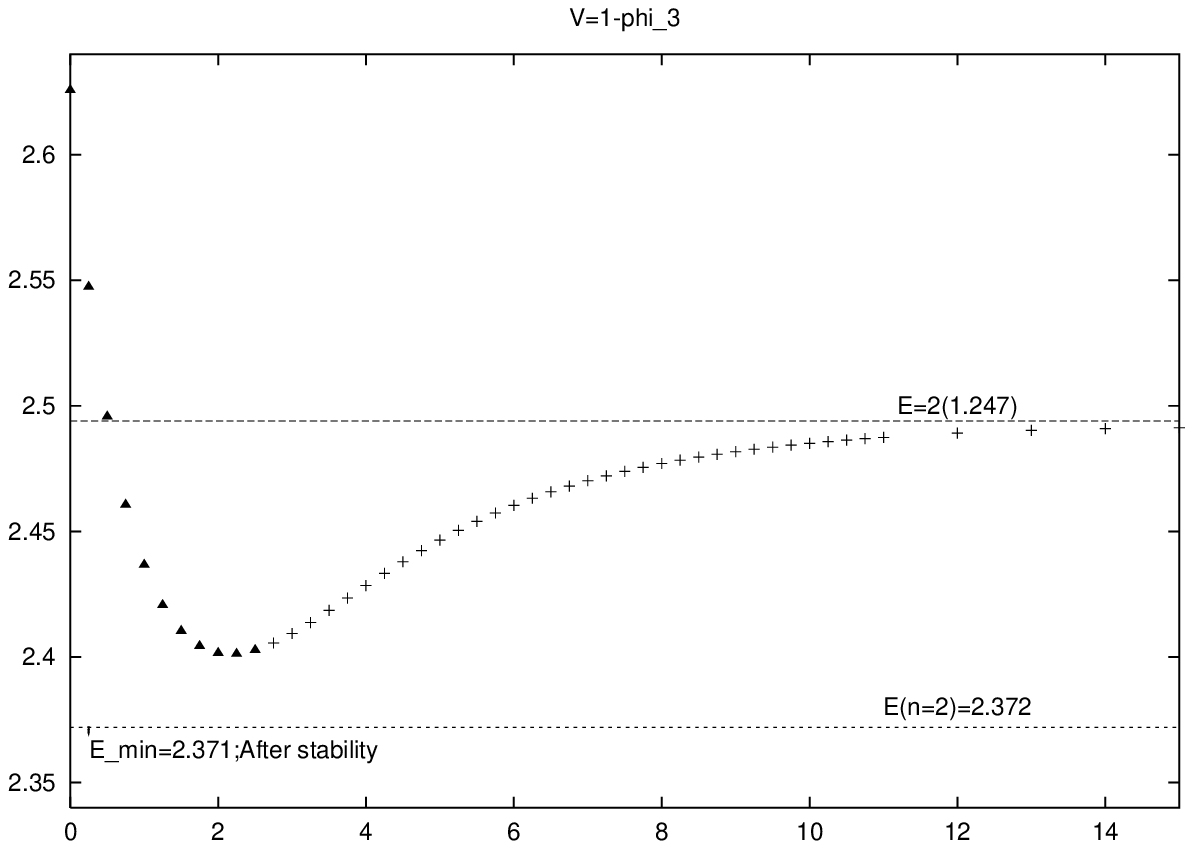}
\includegraphics[width=8cm]{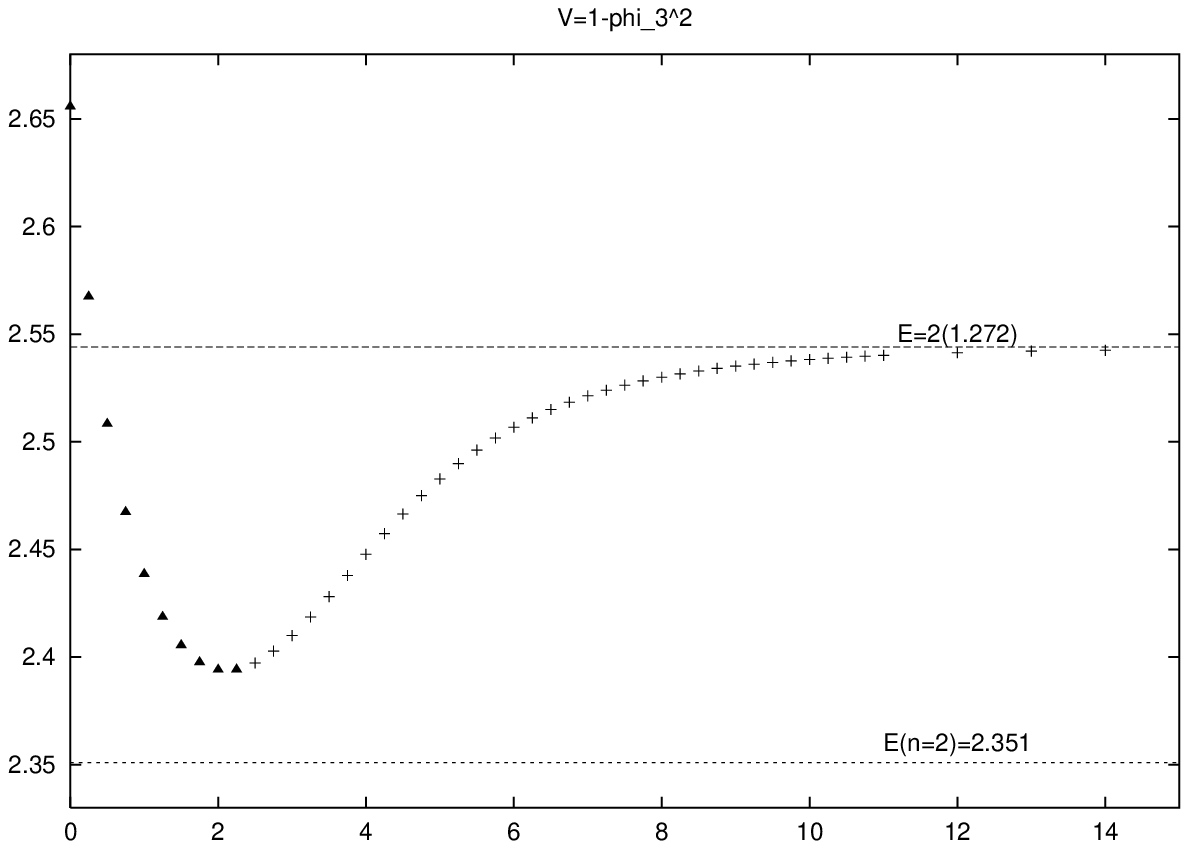}

\includegraphics[width=8cm]{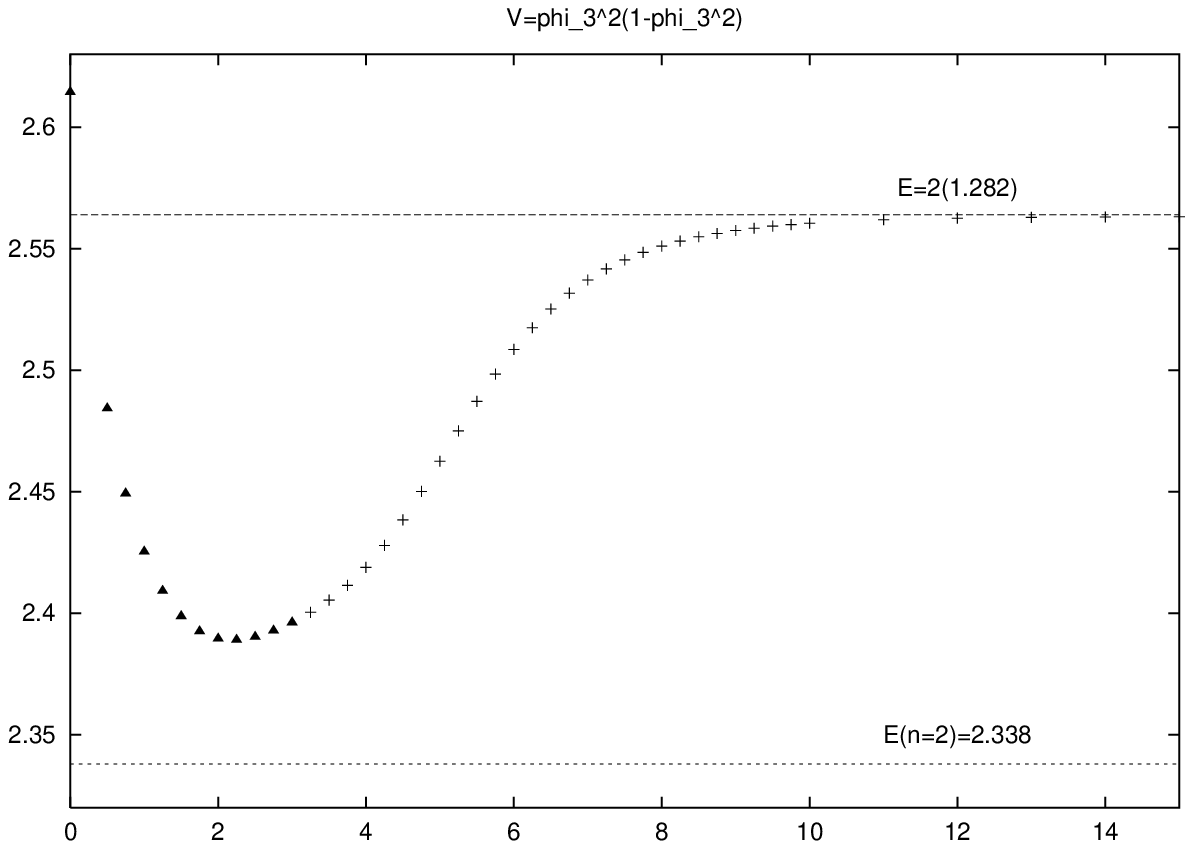}
\includegraphics[width=8cm]{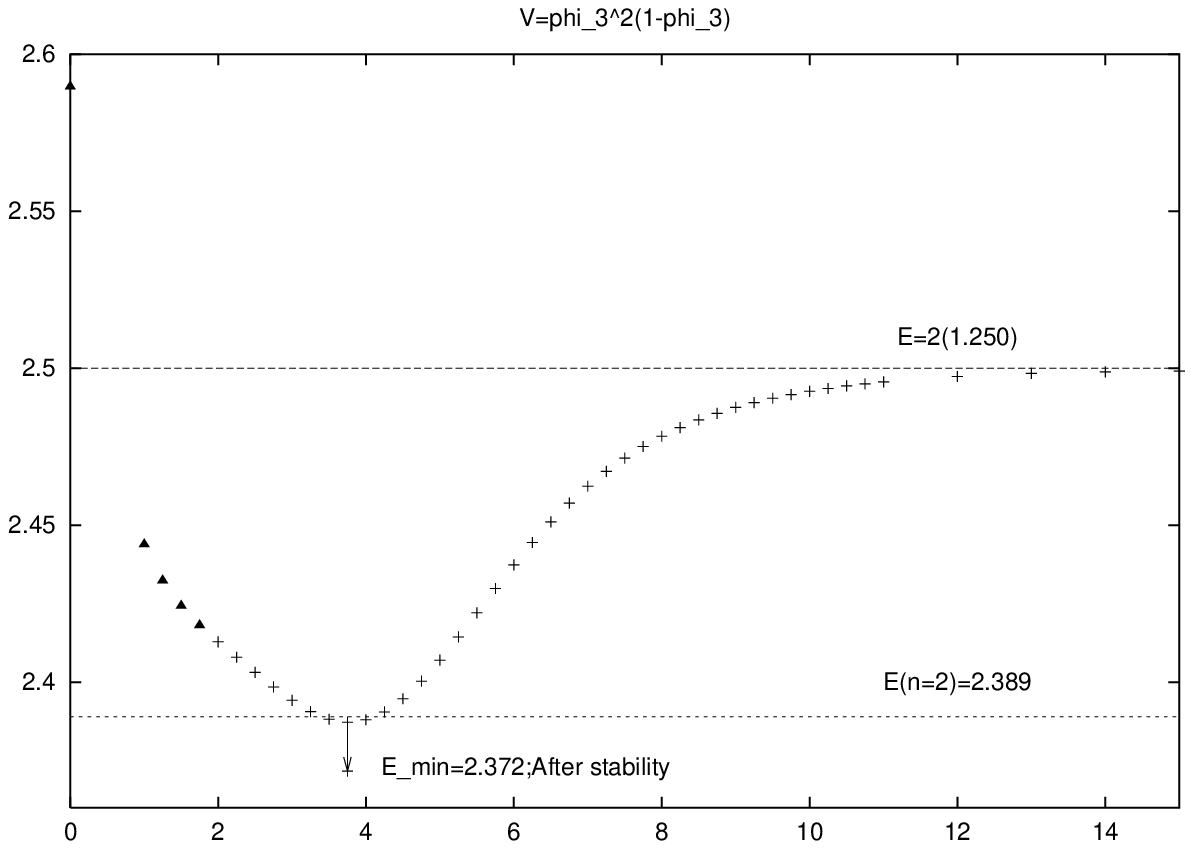}
\caption{The total energy of 2 baby skyrmions as a function of $r$ 
 for different potentials ($G$, defined in (10) given by
respectively,  $G(\phi_3)=1$,
 $G(\phi_3)=(1+\phi_3)$, $G(\phi_3)=(1+\phi_3)\phi_3\sp2$   
 and $G(\phi_3)=\phi_3\sp2$ )}
\end{center}
\end{figure}

%%%%
\section{Further Comments and Conclusions}

The extra coefficient $\phi_{3}^{2}$ causes the potential
to have a further minimum (at $\phi_{3}=0$). The question then arises
of how this change effects the properties of the solutions. 
To perform such a comparison 
of multi-skyrmions of  potentials (\ref{c}) and 
(\ref{b})
we have chosen the parameter $\theta_S$'s in both models to be
such that the 
one skyrmion solutions of these models have the same energy.

Table 4 shows the energy per skyrmion 
for the new baby skyrmion model\cite{weidig:99}
and for the model with the potential (\ref{c}) respectively. 
The results indicate that a further minimum in the potential 
increases the binding of the skyrmions ({\it ie} the multiskyrmions
have lower energies). The same seems to be true for the other two
potentials.

\begin{table}
\begin{center}
\begin{tabular}{|c||c|c|c|c|}\hline
  Charge & Energy per skyrmion& Energy per skyrmion  
  \cr\hline 1 & 1.564 & 1.564 
  \cr\hline 2 & 1.405 & 1.384   
  \cr\hline 3 & 1.371 & 1.344
  \cr\hline 4 & 1.358 & 1.329  
  \cr\hline 5 & 1.352 & 1.322  
  \cr\hline 6 & 1.349 & 1.319
  \cr\hline  
\end{tabular}
\end{center}
\caption{Comparison of energy per skyrmion between 
$V=1-\phi_{3}^{2}$ and $V=\phi_{3}^{2}(1-\phi_{3}^{2})$ }
\end{table}

We have also tried to understand the
difference between the potentials (\ref{d}) and (\ref{a}) (in
the (\ref{a}) the minimum of the potential is very close to $r=0$ while
for (\ref{d}) it is clearly much further out).
A possible suggestion is that, in the latter case, the skyrmions 
move to a distance between them at which $\phi_3=0$. 
Unfortunately this idea is not supported by the results 
of our numerical simulations. 
We have looked at the plots of the energy density
and of the $\phi_3$ at different distances for 2 baby skyrmions
and have found $\phi_3\ne0$.
When we repeated this study for larger $\theta_V$
(to have a more effective potential) the value
of $\phi_3$ decreased but was still nonzero.
So the behaviour of skyrmions is more complicated and it depends 
on the properties of the potential in a more global way.  

We have also calculated the ionisation energies and various break-up 
modes of multiskyrmions for potentials (27) and (28). Comparing 
potentials (25) and (28) we see some further similarity (in each
case the system of 6 skyrmions seems to be the most bound) and in each case, 
eg, the state of 8 skyrmions needs very little energy to break up into (2+6).
This is more pronounced in the case of the potential (28) for which the 
binding is stronger.

In this paper we have looked at baby Skyrme models with more general
potentials. We have found that as the skyrmions involve mappings between
$S\sp2\rightarrow S\sp2$ their properties depend crucially
on whether the potential vanishes at the positions of the 
skyrmions ({\it ie} at $\phi_3=-1$) (we have assumed that at 
spatial infinity $\phi_3=1$). When the potential vanishes 
the skyrmions ``lie on top of each other", when it does not - 
they separate and form interesting lattice like patterns. The
shape of these patterns depends on the details of the potential. 
The same holds for the binding energies of skyrmions in all
models. When the potential is more complicated ({\it ie}
it has further zero) the skyrmions are more bound and in their
patterns are more spread out - however, the actual positions and
distortions depend on the details of the potential.
 
When the skyrmions are spread out the system of skyrmions has many
local minima with some, larger or smaller, potential barriers
between them. Thus for instance, for $V=1-\phi_3$, a system 
of 6 skyrmions has at least 3 local minima, and depending on 
the initial configuration the system can land in any of them.
This is not unexpected and it suggests that similar models in, physically
more relevant, 3 spatial dimensions may also have many local minima.
Thus, the problem of finding multiskyrmion solutions of models
in higher dimensions is clearly very complicated.
\vskip 2cm
{\bf ACKNOWLEDEMENTS}\linebreak
\vskip 0.5cm
We wish to thank Bernard Piette and Tom Weidig for useful comments and
discussions. One of us (PE) thanks the University of Mashhad for a grant
that made her visit to Durham possible and CPT, the University of Durham
for its hospitality.

%%%%%%%%%

\section{Appendix}
Here we make some remarks 
about our numerical procedures 
%%%%%%%%

\paragraph{Profile functions}
We have used the shooting method 
to determine $f(r)$ and have integrated (\ref{profile}) by
a fourth-order Runge-Kutta method for any
 $n$. To avoid a singularity at $r=0$
we have considered $r\simeq\epsilon$ with $\epsilon$ small. 
We have used the formulae
(\ref{f_origin}) and (\ref{fp_origin}) with 10000 lattice
points and with the spacing $dr=0.003$.    
%%%%%%%%%%%%%

\paragraph{Hedgehog static solutions }
In most of our simulations we have used a  $201\times 201$ lattice with 
lattice spacing  $\delta x=\delta y=0.3$.
 For each $(x_i,y_j)$  $f(x_i,y_j)$ was
determined by a linear interpolation using the values
determined by the shooting method.
Given $f$ we calculated $\ph$ using the hedgehog field 
expression (\ref{hedgehog})
%%%%%%%%%%%%%

\paragraph{A linear superposition for static solutions with $n>1$}
In an alternative formulation we have used  a single complex field $W$,
which is related to $\ph$ by 
\begin{equation} \label{relwp}
\phi_1=\frac{W+W^{*}}{1+W W^{*}}\,\,\,\,\,\,\,\,\,
\phi_2=i \frac{W-W^{*}}{1+W W^{*}}\,\,\,\,\,\,\,\,\,
\phi_3=\frac{1-W W^{*}}{1+W W^{*}}\,\,\,\,\,\,\,\,\,
\end{equation}
where $*$ denotes the complex conjugation.
Hence the complex field $W$ when expressed in terms of 
the profile function $f(r)$     
and $\theta$ takes the form 
\begin{equation}
W=\tan\left(\frac{f(r)}{2}\right)e^{-in\theta}.\label{wcom}
\end{equation}
Static initial field configurations with $Q=n$
 were formed by a linear superposition. When the baby skyrmions 
are far from each other (but not too close to the boundary)
we can construct a configuration with charge $n$ from $W$ 
with $n=1$ by a linear superposition
\begin{equation}
W(x,y)=\sum_{\alpha} W_{\alpha}(x-x_{\alpha},y-y_{\alpha}),
\end{equation}
where ($x_{\alpha}$,$y_{\alpha}$) is the location of the centre of the
{$\alpha$}th skyrmion and $W_{\alpha}$ is the field (\ref{wcom})
with $n=1$ and appropriate $f(r)$.
We have used this method 
of constructing initial configurations in some of our simulations.
Namely, we have placed $n$ baby skyrmions
at equal distances from the origin with relative phase shifts 
$\delta\chi=\frac{2\pi}{n}$ between them (for maximal attraction)
and then used  (\ref{relwp}) to get $\ph$.  

%%%%%%%%%

\paragraph{Time evolution of the static solutions}
We integrated the equation of motion for each component of $\ph$ 
independently. In this manner, during the
simulations the
field $\ph$ would gradually move away from the unit sphere ${\cal S}^{2}$. 
To correct this, every few iterations, we kept on rescaling the
field as follows:

\begin{equation}
\phi_{a}\longrightarrow \frac{\phi_{a}}{\sqrt{\ph\cdot\ph}} 
\end{equation}

and

\begin{equation}
\partial_{t}\phi_{a}\longrightarrow\partial_{t}\phi_{a}-
\frac{\partial_{t}\ph\cdot\ph}{\ph\cdot\ph}\phi_{a}.
\end{equation}

Another problem we have had to face involved using a finite lattice.
Thus we had to make sure that the boundary effects
did not alter our results. To be certain of this we varied 
lattice spacing and the number of lattice points.

When the fields were not at the minimum of the energy we allowed them
to flow to this minimum - reducing the energy by a damping term.

\begin{equation}
\partial_{tt}{\phi}_{a}=K_{ab}^{-1} {\cal
  F}_b\left(\ph,\partial_{t}\ph,\partial_{i}\ph\right)
  -\gamma\partial_{t}\phi_a,
\end{equation}
where $\gamma$ is the damping coefficient. We set $\gamma$ to 0.1.
This term takes the energy out of system.

We have used a fourth-order Runge-Kutta method to simulate the
evolution of the field in time working in double precision.
The time step was $\delta t=0.1$ throughout. We have 
performed our simulation using the $\ph$ formulation 
with the derivatives replaced by finite differences\footnote{we have used
the 9-point laplacian}(as explained in \cite{pz:98}, see also \cite{num}).

\bibliographystyle{plain} \bibliography{rasol_ref}

\begin{thebibliography}{1}

\bibitem{lpz:90}
R~A Leese, M~Peyrard, and W~J Zakrzewski.
\newblock Soliton scattering in some relativistic models in (2+1) dimensions.
\newblock {\em Nonlinearity}, 3:773--807, 1990.

\bibitem{psz:95b}
B~M A~G Piette, B~J Schoers, and W~J Zakrzewski.
\newblock Dynamics of baby skyrmions.
\newblock {\em Nuclear Physics B}, 439:205--235, 1995.

\bibitem{psz:95a}
B~M A~G Piette, B~J Schoers, and W~J Zakrzewski.
\newblock Multisolitons in two-dimensional skyrme model.
\newblock {\em Z. Phys. C}, 65:165--174, 1995.

\bibitem{pz:93}
B~M A~G Piette and W~J Zakrzewski.
\newblock Skyrmion dynamics in (2+1) dimension.
\newblock {\em Chaos, Solitons and Fractals}, 5:2495--2508, 1995.

\bibitem{pz:98}
B~M A~G Piette and W~J Zakrzewski.
\newblock Numerical integration of (2+1)d pdes for $s^2$ valued functions.
\newblock {\em J. Comp. Phys.}, 145:359--381, 1998.

\bibitem{num}
W~H Press, S~A Teukolsky, W~T Vetterling, and B~P Flannery.
\newblock {\em Numerical Recipes in C}.
\newblock Cambridge University Press, 1992.

\bibitem{sky:61}
T~H~R Skyrme.
\newblock A nonlinear field theory.
\newblock {\em Proc. Roy. Soc. A}, 260:127--138, 1961.

\bibitem{sut:91}
P~Sutcliffe.
\newblock The interaction of skyrme-like lumps in (2+1) dimensions.
\newblock {\em Nonlinearity}, 4:1109--1121, 1991.

\bibitem{weidig:99}
T~Weidig.
\newblock The baby skyrme models and their multi-skyrmions.
\newblock {\em Nonlinearity}, 12:1489--1503, 1999.

\end{thebibliography}
\end{document}